\documentstyle[aps,prb,epsf]{revtex}

\newcommand{\be}{\begin{equation}}
\newcommand{\ee}{\end{equation}}
\newcommand{\bfig}{\begin{figure}}
\newcommand{\efig}{\end{figure}}
\newcommand{\bea}{\begin{eqnarray}}
\newcommand{\eea}{\end{eqnarray}}
\newcommand{\ba}{\begin{array}}
\newcommand{\ea}{\end{array}}

\begin{document}
\title{Statics and Dynamics of Phase Segregation in Multicomponent Fermion Gas}
\author{K. Esfarjani\cite{perk}, S.T. Chui\cite{perc}, V. Kumar\cite{perv}, and Y. Kawazoe}
\address{Institute for Materials Research, Tohoku University, Sendai 980-8577, Japan\\
}
\maketitle
\date{\today}

\begin{abstract}
We investigate the statics and dynamics of spatial phase segregation process
of a mixture of fermion atoms in a harmonic trap using the density
functional theory. The kinetic energy of the fermion gas is written in terms
of the density and its gradients. Several cases have been studied by
neglecting the gradient terms (the Thomas-Fermi limit) which are then
compared with the Monte-Carlo results using the full gradient corrected
kinetic energy. A linear instability analysis has been performed using the
random-phase approximation. Near the onset of instability, the fastest
unstable mode for spinodal decomposition is found to occur at $q=0$.
However, in the strong coupling limit, many more modes with $q\approx K_F$
decay with comparable time scales.

\noindent
\end{abstract}

\pacs{PACS${\#}$:  03.75.Fi; 64.75.+g   }

\widetext

\section{Introduction}

Recent realizations of two\cite{jila1,jila2} and three\cite{mit} component
alkali Bose-Einstein condensates (BEC's) in a trap provide us with new
systems to explore the physics in otherwise unachievable parameter regimes.
\cite{law,ho,cta}. Dramatic results have recently been observed in the phase
segregation dynamics of mixtures of Rb\cite{jila1,jila2} and Na\cite{mit}
gases. Periodic spatial structures were found at intermediate times which
then recombine at a later time.

Phase segregation phenomena have been much studied in materials science and
these can be understood using {\it classical mechanics}. Spatial modulations
have also been observed, for example, in AlNiCo alloys.\cite{iwama} These
were explained in terms of a concept called spinodal decomposition.\cite
{cahn} When a system is quenched from the homogeneous phase into a
broken-symmetry phase, the ordered phase does not order instantaneously.
Instead, different length scales set in as the domains form and grow with
time. For the BEC's, however, {\it quantum mechanics} play an important
role. It has been shown\cite{Ao} that it is possible to have an analogous
spinodal decomposition, which manifests some of the phenomenology including
a periodic spatial structure at an intermediate time that is now determined
by quantum mechanics. The time scale provides for a self-consistent check of
the theory and is consistent with the experimental results.\cite{jila2} The
growth of domains at later times is now determined by quantum tunneling and
not by classical diffusion.

Recently, it became possible\cite{demarko} to cool a single component system
of about a million $^{40}$K fermionic atoms in a magnetic trap below the
Fermi temperature, $T_{F}$, leading to the realization of a spin-polarized
fermion gas of atoms. Similar to electrons in a solid, the dilute gas of
atoms fills all the lowest energy states below the Fermi energy, $E_{F}$.
The transition to this quantum degenerate state is gradual as compared to
the abrupt phase transition into a Bose condensate. For single component
fermionic systems, however, the equilibrium is difficult to achieve as the $s
$-wave elastic collisions are prohibited due to Pauli exclusion principle.
In the experiments of DeMarco and Jin\cite{demarko}, this was circumvented
by using a mixture of two nuclear spin states of $^{40}$K atoms for which $s$%
-wave collisions are allowed. One of the manifestations of quantum mechanics
was the nature of momentum distribution which differed from the well known
classical gaussian distribution. This system corresponds to the weak
coupling limit in which the physical properties are close to those of a
non-interacting fermion gas. The other system which is being explored\cite
{li} is the gas of $^{6}$Li atoms. 
Mixtures of fermions interating with the Coulomb interaction have been
studied in the context of the electron-hole fluids\cite{BRAC}. For fermions
mixtures on a lattice site interacting with the Hubbard Hamiltonian, the
{\bf partial} phase segregation leads to antiferromagnetism.
Thermodynamic properties as well as density and momentum distributions of
spin-polarized fermionic gas of atoms in a harmonic trap have been studied
in recent years\cite{rok,bruun,sch}. Butts and Rokhsar\cite{rok} have
obtained universal forms of the spatial and momentum distributions for a
single component spin-polarized non-interacting fermion gas using the
Thomas-Fermi (TF) approximation, whereas Schneider and Wallis\cite{sch} have
studied the effects of shell closure for small number of atoms, similar to
the nuclear shell model. Bruun and Burnett\cite{bruun} have studied an
interacting fermion gas of $^{6}$Li atoms which have a large negative
scattering length. Such an interaction could also lead to the possibility of
superfluid state\cite{hou} in these systems. In the present paper, we
consider mixtures of these new finite systems of ultracold fermionic atoms
with a positive scattering length in the limit of both weak and strong
coupling and explore the equilibrium and non-equilibrium quantum statistical
physics using the TF approximation, Monte Carlo simulations, and the random
phase approximation.

In section II we present the equilibrium static properties of mixtures of
fermionic atoms in different parameters regimes using both the TF and the
Monte Carlo simulations. In section III, we study the dynamics of phase
segregation of such mixtures using a linear stability analysis. Finally,
conclusions will be presented in section IV.

\section{Statics}

We first start with the statics of a two component fermion gas of atoms with
masses $m_1$ and $m_2$ and particle numbers $N_1$ and $N_2$. This is assumed
to be confined in an azimuthally symmetric harmonic trap with radial and
axial frequencies $\omega$ and $\lambda\omega$, respectively which are
considered to be the same for both the components. Unlike the electron gas
in matter, the fermion gas of atoms is neutral and dilute. Therefore,
significant interactions between atoms are only short-ranged and that would
be responsible for any phase segregation in the system. In the long
wavelength limit, the system can be well described by the density functional
theory and the total energy can be written as

\begin{equation}
E=\int [\sum_{\sigma} E_{0\sigma}(\rho_{\sigma})+g\rho_1({\bf r})\rho_2({\bf %
r})] {\rm d}{\bf r}.
\end{equation}
Here $E_{0\sigma} =\frac{\hbar ^2}{2m_{\sigma}}\tau_{\sigma}({\bf r})+{\frac{%
1 }{2}} m_{\sigma}\omega^2(x^2+y^2+ \lambda^2z^2)\rho_{\sigma}({\bf r}) $ is
the non-interacting part of the energy density and $\rho _{\sigma}({\bf r})$
is the particle density of the component $\sigma=1,2$ with $\int
\rho_{\sigma}({\bf r})d{\bf r} = N_{\sigma}$. The interaction term has been
approximated by the contact potential $g\delta({\bf r}-{\bf r}^{\prime})$. $g
$ is related to the scattering length $a$ by $g=2\pi\hbar^2 a/{\bar m}$,
with $\bar m = m_1m_2/(m_1+m_2)$. In accordance with the experiments, we
take $a$ to be positive and consider only the $s$-wave scattering.
Therefore, the contribution to the interaction term is non-zero only when
the species are different or are in different hyperfine states as in
experiments. From the Pauli exclusion principle, there is no contact
interaction between particles of the same species (spin). In a more general
treatment including $p$-wave scattering there would be additional terms
involving interaction between identical species also. But these are small,
and thus neglected.

For the kinetic energy density $\tau_{\sigma}$ we use a local approximation
including the first and second derivatives of the particle density,
\begin{equation}
\tau_{\sigma} ({\bf r})=\frac 35(6\pi ^2)^{2/3}\rho _{\sigma}({\bf r})^{5/3}
+\frac 1{36}\frac{| \nabla \rho _{\sigma}({\bf r}) | ^2}{\rho _{\sigma}({\bf %
r})}+\frac 13\nabla ^2\rho _{\sigma}({\bf r}).
\end{equation}

The first term represents the Thomas-Fermi (TF) approximation to the kinetic
energy. The second term is ${\frac{1}{9}}| \nabla \sqrt{\rho_{\sigma}} \,|^2$
and represents the gradient correction to the kinetic energy. The integral
of the third term extended to infinity vanishes, and thus it will not be
included in the calculations. The Monte-Carlo results confirm that the
gradient term is at least 2 orders of magnitude smaller than the TF term,
but this term is important in that it can break the symmetry of the ground
state and lead to asymmetric states with a lower energy .

Without the interaction term in (1), the system behaves in the same fashion
as the one component system for which Butts and Rokhsar\cite{rok} obtained $%
E_F$ to be related to the total particle number $N$ by $E_F=\hbar
\omega(6\lambda N)^{1/3}$. Defining $R_F=(2E_F/m\omega^2)^{1/2}$ (giving the
characteristic size of the gas), and $K_F=(2mE_F/\hbar^2)^{1/2}$ (momentum
of a free particle of energy $E_F$), they calculated the density profile at
T=0 to be given by
\begin{equation}
\rho_{{\rm non-interacting}}({\bf r})=\rho_0 \left[1-{\bar r}^2/R_F^2 \right]%
^{3/2},  \label{rho0}
\end{equation}
with ${\bar r}^2=x^2+y^2+\lambda^2z^2$, $\rho_0=8 N\lambda / \pi^2 R_F^3=
K_F^3/6 \pi^2$. In the TF approximation, the trapping potential can be
treated to be locally constant and we can define a local Fermi wavevector, $%
k_F({\bf r})$ so that $E_F = \hbar^2 k_F^2({\bf r})/2m +V({\bf r})$, and the
density at T = 0 can also be written as $\rho_{{\rm non-interacting}}({\bf r}%
) = k_F^3({\bf r})/6 \pi^2$.

We now examine the properties of the mixed (two-component)
interacting system and will show how the repulsive interaction
modifies this non-interacting density profile as well as other
properties of the system. The strength of the coupling, which
controls the phase segregation, depends on the dimensionless
parameter which is the ratio between the interaction and the
kinetic energies, namely $g\rho_1\rho_2/[{\frac{3\hbar^2}{{10}}}(6
\pi^2)^{2/3} ({\rho_1^{5/3}/m_1+\rho_2^{5/3}/m_2})] $.
In the simple case of equal masses ($m_1 = m_2 = m$) and densities ($\rho_1
= \rho_2 = \rho$) of the two components, this simply scales as $a K_F$. This
means that the coupling would be stronger if $a$ or the density is large.
Also as $E_F$ is proportional to the frequency of the trap at constant $N$
(a higher frequency leads to a larger separation between the levels), the
coupling would be large for higher frequencies. From now on, to measure the
strength of the interaction, we will use the dimensionless parameters $%
c_{\sigma} = K_{F\sigma} \,a /\pi$, where $K_{F\sigma} = (2 m_{\sigma}
\mu_{\sigma})^{1/2} /\hbar $ , or in the case of equal chemical potentials,
just $c = K_F \,a /\pi$.

For a general two-component system with chemical potentials $\mu_1$ and $%
\mu_2$, the ground state is obtained by minimizing the thermodynamic
potential $\Omega= E-\int (\mu_1 \rho_1+\mu_2 \rho_2) {\rm d}{\bf r} $. This
leads to the following system of equations:\newline
\begin{eqnarray}
\frac{\partial \Omega}{\partial \rho_1({\bf r})}& =& \frac{\hbar ^2}{2m_1} %
\left[(6 \pi^2 \rho_1)^{\frac{2}{3}} - {\frac{1}{36}}( \left| \frac{\nabla
\rho_1 } {\rho_1} \right|^2 + 2 \frac{\nabla^2 \rho_1}{\rho_1}) \right] +({%
\frac{1 }{2}} m_1 \omega^2 {\bar r}^2 - \mu_1+ g \rho_2) =0 \\
\frac{\partial \Omega}{\partial \rho_2({\bf r})} &=& \frac{\hbar ^2}{2m_2} %
\left[(6 \pi^2 \rho_2)^{\frac{2}{3}} - {\frac{1}{36}}( \left| \frac{\nabla
\rho_2 } {\rho_2} \right|^2 + 2 \frac{\nabla^2 \rho_2}{\rho_2}) \right] +({%
\frac{1 }{2}} m_2 \omega^2 {\bar r}^2 - \mu_2+ g\rho_1) =0.
\end{eqnarray}

Similar to the one-component case, one can rewrite the above in a
dimensionless form by introducing for each of the species $\sigma $, the
following quantities: $R_{\sigma }=[2\mu _{\sigma }/m_{\sigma }\omega ^{2}]^{%
\frac{1}{2}}$, $\rho _{\sigma 0}=K_{F\sigma }^{3}/6\pi ^{2}$, ${\cal G}%
_{\sigma }=g\rho _{{\bar{\sigma}}0}/\mu _{\sigma }$, and $n_{\sigma }({\bf r}%
)=\rho _{\sigma }({\bf r})/\rho _{\sigma 0}$. Here ${\bar{\sigma}}$ = 3-$%
\sigma $. If one neglects the smaller terms containing derivatives of $\rho $
(the TF limit), one obtains the following algebraic equations satisfied by
the dimensionless densities $n_{1}$ and $n_{2}$ for any coupling strength $%
{\cal G}_{\sigma }$:
\begin{eqnarray}
n_{1}^{2/3} &=&1-{\bar{r}}^{2}/R_{1}^{2}-{\cal G}_{1}n_{2}  \nonumber \\
n_{2}^{2/3} &=&1-{\bar{r}}^{2}/R_{2}^{2}-{\cal G}_{2}n_{1}.
\label{densities}
\end{eqnarray}
We see that the effect of the additional ${\cal G}_{\sigma }n_{{\bar{\sigma}}%
}$ term, i.e. the interaction, is to deplete the regions where $n_{{\bar{%
\sigma}}}$ is highest (without necessarily leading to a phase segregation).

When there is phase segregation, the interface energy is proportional to the
square root of the coefficient of the gradient term\cite{ac} and it often
serves to distinguish different configurations. In that case, their effect
cannot be neglected and these are included in the Monte Carlo simulations.
We next discuss some special cases in the TF limit.

\subsection{TF limit: Similar densities: ($\protect\mu_1=\protect\mu_2$) for
any coupling}

To simplify the notations, we will use: $\mu_1=\mu_2=\mu;\ R_1=R_2=R;\ {\cal %
G}_1={\cal G}_2={\cal G}$. In this case, three solutions to Eq. (6) will
correspond to $n_1=n_2$, of which only one is physical with $n_1> 0$. If a
solution $n_2=f(n_1)$ exists, by symmetry, the other one is necessarily $%
n_1=f(n_2)$. These solutions with $n_1\neq n_2$ can be obtained numerically.
The real solutions are plotted in Fig. \ref{thermo}, where the $n_1=n_2$
solution is referred to as ``Sym", and the other conjugate (asymmetric)
solutions are referred to as ``A1" and ``A2". Below we discuss these
solutions in the weak and strong coupling limits.

\subsubsection{ Weak or intermediate coupling regime}

In this case we look for symmetric solutions ($n_1=n_2=n$). Equation (\ref
{densities}) then reduces to (dropping the subscripts):
\begin{equation}
n({\bf r})^{2/3}=1-{\bar r}^2/R^2-{\cal G} \,n({\bf r}),  \label{sym}
\end{equation}
which can be solved easily numerically to give the density profile of the
non-segregated phase. It is possible to show that after proper rescaling,
the result for all coupling strengths and at any point can be summarized in
a single universal curve in Fig. \ref{thermo}. If $n({\bf r})$ is a solution
to Eq. (\ref{sym}), then ${\cal N}=n \,{\cal G}^3$ versus ${\cal P}=\left[1-{%
\bar r}^2/R^2 \right] {\cal G}^2$ is the universal function of Fig. \ref
{thermo} satisfying ${\cal N}^{2/3} + {\cal N} -{\cal P} =0$. For small
couplings and near the boundary (${\cal P} \approx 0; \, {\cal N}^{2/3} \gg
{\cal N} \, \Leftrightarrow {\cal N}={\cal P}^{3/2}$ ), this curve is a
power law and in fact tends to the non-interacting density $n({\bf r})
\approx \left[1-(x^2+y^2+\lambda^2z^2)/R^2 \right]^{3/2}$.

\subsubsection{Strong coupling regime}

The above situation, however, can not be always sustained. In the strong
coupling limit, we can have phase segregation ($n_1 \ne n_2$), and one needs
to go back to Eq. (\ref{densities}) which now admits lower energy solutions
that are not ``permutation symmetric'':
\begin{eqnarray}
n_1^{2/3} + {\cal G} n_2 &=& 1-(x^2+y^2+\lambda^2z^2)/R^2 \Leftrightarrow
{\cal N}_1^{2/3} + {\cal N}_2 = {\cal P} \Leftrightarrow {\cal N}_1^2 = (%
{\cal P}-{\cal N}_2)^3  \nonumber \\
n_2^{2/3} + {\cal G} n_1 &=& 1-(x^2+y^2+\lambda^2z^2)/R^2 \Leftrightarrow
{\cal N}_2^{2/3} + {\cal N}_1 = {\cal P} \Leftrightarrow {\cal N}_2^2 = (%
{\cal P}-{\cal N}_1)^3,  \label{strong}
\end{eqnarray}
where we used the same simplifying notations as before.
As previously mentioned, the symmetric solution ${\cal N}_1 = {\cal N}_2$
always exists. This can be exploited to reduce the above equations to a
quadratic equation, which is analytically more transparent.

Subtracting the above equations from each other and dividing out by ${\cal N}%
_1-{\cal N}_2$, we obtain,
\begin{equation}
{\cal N}_1+{\cal N}_2 = ({\cal P}-{\cal N}_2)^2+ ({\cal P}-{\cal N}_1)^2 +(%
{\cal P}-{\cal N}_2)({\cal P}-{\cal N}_1).
\end{equation}
This quadratic equation can be solved for ${\cal N}_1$ in terms of ${\cal N}%
_2$.

The solutions will all be axially symmetric in that they are functions of ${%
\bar{r}}^{2}$ only. In actuality, the axial symmetry can also be broken, but
we do not find it here since we neglected the terms in gradient of the
particle density in the kinetic energy. The broken symmetry solutions will
be discussed in the subsection E where we present results obtained from the
Monte Carlo simulations incorporating these terms. In Fig. \ref{thermo}, the
solutions with $n_{1}\neq n_{2}$ can be seen in the limit of small reduced
distance and large ${\cal P}$. The bifurcation point where these solutions
start to occur, corresponds, from numerical results, to ${\cal P}_{c}\approx
0.741$, and ${\cal N}_{c}=n{\cal G}^{3}\approx 0.296$, which separate the
strong coupling regime from the weak one. In both figures, the symmetric
solution is drawn with solid line, and the asymmetric ones with dashed
lines. Actually, at the bifurcation point, we have exactly ${\cal G}n^{\frac{%
1}{3}}=2/3$ as will be shown in the TF linear stability analysis section
below. Since ${\cal G}^{2}={\cal P}/(1-{\bar{r}}^{2}/R^{2})\geq {\cal P}$,
the smallest coupling ${\cal G}_{c}$ for the unequal solutions to occur
satisfies ${\cal G}_{c}=\sqrt{{\cal P}_{c}}$. Since ${\cal G}%
=(4/3)K_{F}a/\pi $, we find a critical dimensionless coupling $c=(K_{F}a/\pi
)_{c}\approx 0.646$. We shall come back and compare this value with that
obtained with a different approach.

\subsection{TF limit: Very different densities: ($\protect\mu_1 \gg \protect%
\mu_2$) for any coupling}

One can also treat the case where one of the species is a minority ($\mu_1
\gg \mu_2$). If we assume $\mu_1 = \lambda^2 \mu_2$, then $R_1 = \lambda
R_2; \, K_F{}_1 = \lambda K_F{}_2; \rho_{10} = \lambda^3 \rho_{20};\, {\cal G%
}_2 = \lambda^5 {\cal G}_1$, and $n_{\sigma} \sim 1$. The density
distribution of the majority species will be weakly perturbed. Referring to
Eqs. (\ref{densities}), one can see that the coupling ${\cal G}_1 = g
\rho_{20} /\mu_1 $ becomes very small and maybe neglected. Thus a good
approximation is to assume $\rho_1\approx\rho_{{\rm non-interacting}}$. The $%
{\cal G}_2$ term in the second equation, however, is a large quantity, and
will strongly affect the particle density $n_2$. Therefore,
\begin{equation}
n_2({\bf r}) \approx \left[ 1- {\bar r}^2/R_2^2 - {\cal G}_2 [ 1- {\bar r}%
^2/R_1^2 ]^{\frac{3}{2}}\right]^{\frac{3}{2}}.
\end{equation}
In the presence of the majority species, the number of atoms of minority
species will be much less than their non-interacting counterparts with the
same chemical potential. As we can see from the above equation, their
number, even at the origin is reduced by a factor of $(1-{\cal G}_2)^{\frac{3%
}{2}}$. We find that for a large enough ${\cal G}_2$ the density ${\cal N}_2$
is depleted from the center (see also Fig. \ref{thermo}b, curve A2).

\subsection{TF limit: linear instability analysis}

We next study the fluctuations of the system about its equilibrium
configuration in the TF limit by expanding the thermodynamic potential $%
\Omega$ upto second order in the particle density variation $\delta \rho$
about its minimum which was computed above. The sign of the second
derivative of $\Omega$ will decide the stability of the symmetric phase. A
phase segregation occurs when the Hessian (second derivative matrix) ceases
to be positive definite. If the transition is first order, it would have
already occurred before reaching a negative second derivative.
The second derivative from Eqs. (3) and (4) is just a $2 \times 2$ matrix:
\begin{equation}
\frac{\partial^2 \Omega}{\partial \rho_{\sigma} \partial
\rho_{\sigma^{\prime}}} = \frac{\hbar ^2}{2m_{\sigma}} {\frac{2}{3}} (6
\pi^2)^{\frac{2}{3}} \, \rho_{\sigma}^{-{\frac{1 }{3}}} \, \delta_{\sigma
\sigma^{\prime}} + g \,(1-\delta_{\sigma \sigma^{\prime}}) .
\end{equation}
The phase instability criterion thus becomes $\omega_-=0$ where $\omega_-$
is the smallest eigenvalue of the Hessian matrix; implying:
\begin{equation}
\frac{\hbar ^2}{2\sqrt{m_1 m_2}} {\frac{2}{3}} (6 \pi^2)^{\frac{2}{3}} \,
(\rho_1 \rho_2)^{-{\frac{1 }{6}}} = g \Leftrightarrow {\frac{\mu }{\rho_0}} {%
\frac{2 }{3}} n^{-1/3} = g \,\,{\rm if }(\rho_1=\rho_2)
\end{equation}
Thus, in the symmetric case ($\mu_1=\mu_2; \, \rho_1=\rho_2$), the
instability will first occur locally at the point where the relation ${\cal N%
}^{1/3}={\cal G} n^{\frac{1}{3}} = 2/3 $ is satisfied. This implies that $%
{\cal N}=0.296$, which is exactly the critical ${\cal N}_c$ obtained earlier
from a different analysis. These two instabilities occuring at the same
point suggest that, within the adopted model (TF), the transition might be
of second order.

\subsection{Possibility of density modulation instability}

Similar to the electron gas which has several kinds of instabilities such as
ferromagnetism, antiferromagnetism, charge density wave, superconductivity,
etc... these two-component systems might also exhibit other types of
instabilities. To investigate them, we will assume the homogeneous case ($%
\omega=0$) as the analysis can be made simpler by using the Fourier
decomposition of the density. To get some understanding of the nonuniform
systems (such as in a trap), one can assume in a semiclassical
approximation, that the Fermi momentum depends on the position, as before.

The density for the species $\sigma$ can be written as the sum of its
Fourier components: $\rho_{\sigma}({\bf r}) = {\bar{\rho}}_{\sigma} + \sum_{%
{\bf q} \ne 0} \,\rho_{\sigma {\bf q}} \, e^{i {\bf q} .{\bf r}}$, with ${%
\bar{\rho}}_{\sigma} \gg \rho_{\sigma {\bf q}}$. Substituting this
expression in the thermodynamic potential $\Omega$, expanding up to second
powers of $\rho_{\sigma {\bf q}}$, and minimizing $\Omega$ with respect to
the Fourier components, we obtain:
\begin{equation}
\frac{\partial \Omega}{\partial \rho_{\sigma {\bf q}}} = \frac{\hbar ^2}{%
2m_{\sigma}} \left[ {\frac{2 }{3}} (6 \pi^2 {\bar{\rho}}_{\sigma})^{\frac{2}{%
3}} \, \frac{\rho_{\sigma -{\bf q}}}{{\bar{\rho}}_{\sigma} } + \frac{1}{36 }
q^2 \frac{\rho_{\sigma -{\bf q}}}{{\bar{\rho}}_{\sigma}} \right] + g \rho_{{%
\bar{\sigma}} -{\bf q}} =0
\end{equation}
\begin{equation}
\frac{\partial \Omega}{\partial {\bar{\rho}}_{\sigma} } = \frac{\hbar ^2}{%
2m_{\sigma}} (6 \pi^2 {\bar{\rho}}_{\sigma})^{\frac{2}{3}} -\mu_{\sigma} + g
{\bar \rho}_{{\bar{\sigma}}} =0
\end{equation}
Assuming $6 \pi^2 {\bar{\rho}}_{\sigma} = {\bar{k}}_{\sigma}^3$ (note that
in the presence of interactions, the average density and Fermi momentum,
which we denote here by ${\bar{\rho}}_{\sigma}$ and ${\bar{k}}_{\sigma}$
respectively, are different from their non-interacting values),
the above equations are simplified to:
\begin{eqnarray}
(1+ \frac{q^2 }{24 {\bar{k}}_{\sigma}^2}) \rho_{\sigma {\bf q}}+ 2\,({\frac{{%
\bar{k}}_{\sigma} a }{\pi}})\,\rho_{{\bar{\sigma}} {\bf q}} &=& 0 \\
\frac{\hbar ^2 {\bar{k}}_{\sigma}^2}{2m}+ {\frac{4 }{3}}\,({\frac{{\bar{k}}_{%
\bar \sigma} a }{\pi}}) \,\frac{\hbar ^2 {\bar{k}}_{\bar \sigma}^2}{2m} =%
\frac{\hbar ^2 K_{F\sigma}^2}{2m} &=& \mu_{\sigma}  \label{equil}
\end{eqnarray}

It is clear from the above equations that if $a=0$ then $\rho_{\sigma {\bf q}%
}=0$ is a solution (uniform density if no coupling). For $a>0$, we have $%
\rho_{\sigma {\bf q}}$ and $\rho_{{\bar{\sigma}} {\bf q}}$ of opposite signs
for all ${\bf q}$. This means that there is phase segregation for repulsive
couplings. Furthermore,
if $a<0$, there will be density modulation in the small ${\bf q}$ limit (the
functional we considered is valid in the long wavelength limit). We shall
return to this point in section III where the dynamics are treated.

One can also note that the transition points of Eq. \ref{equil} and
previously studied Eq. \ref{densities} are the same (in the $\omega =0$ and $%
\mu _{1}=\mu _{2}$ case), since they are derived from the same functional.
Indeed from the positive-definiteness of the functional $\Omega $ in this
representation, one obtains that the transition occurs for ${{\bar{k}}%
_{\sigma }a/\pi }=1/2$. Inserting this critical value into Eq. \ref{equil},
one finds the relation between the non-interacting Fermi wavevector $K_{F}$
and the interacting one ${\bar{k}}_{\sigma }$ at the transition point: $%
K_{F}={\bar{k}}_{\sigma }\sqrt{5/3}$ which then implies
\begin{equation}
c=K_{F}a/\pi ={\frac{1}{2}}\sqrt{\frac{5}{3}}\approx 0.645,
\label{ccritical}
\end{equation}
which is exactly the same value as obtained from the numerical result of the
previous section.

\subsection{General case: Monte Carlo results}

The density distribution that extremizes the energy functional in Eq. (1)
can be obtained by a Monte Carlo simulation with a weighting factor exp$%
(-E/T)$ for a parameter T that is sufficiently low. This is basically the
simulated annealing method and has been exploited successfully in earlier
treatment\cite{cta} of the corresponding Bose system described by a
Gross-Pitaevski functional.

We approximate the volume integral of the energy functional by a discrete
sum. Using the scaled radius $\bar r$, we sample a lattice inside a sphere
of diameter $2R$ consisting of 40 sites along the diameter, making a total
of 33398 sites. The derivative term is approximated by a finite difference.
For simplicity, we show here results for the case when the two components
have the same mass.

We first show in Fig. \ref{mc_charge} the density profile of component 1 as
a function of x and y for z=0 for the weak coupling case with no phase
segregation. The values of different parameters were chosen to be $%
\omega=135 \times 2 \pi \,{\rm rad/sec}$, $a=135\, a_{{\rm Bohr}}$, $%
\lambda=0.14$, and $N_1=N_2 =10^6$ ($\mu_1=\mu_2=1.626\times 10^{-29} J$);
roughly corresponding to the experimental parameters of the $^{40}K$ system
\cite{demarko}. In these experiments, we estimate $c=K_F a/\pi = 0.032$, $%
R_F=26 \,\mu$m, and ${\cal G}=0.042$. The density profile for component 2 is
the same and hence is not shown.

In the limit of strong interaction, phase segregation starts, and as
mentioned earlier, the system can now also break cylindrical symmetry. This
happens when $K_Fa$ is large enough, which in turn can be achieved with only
large $K_F$, only large $a$, or both. To illustrate this, we show in Fig.
\ref{mc_charge1} 
the density profiles for components 1 and 2 for the case of only large $a$
with $a=30000 \,a_{{\rm Bohr}},$ $\mu_1=\mu_2=1.86\times 10^{-30}J,$ and $%
\omega=300 \,{\rm rad/sec}$. In this case, $c=K_F a/\pi=2.39$, $R_F= \,25 \mu
$m, and ${\cal G}=3.19$.

For the case of both large $K_F$ and $a$, we show in Fig. \ref{mc_chargea}
the density profiles for $a=3000 \,a_{{\rm Bohr}},$ $\mu_1=\mu_2=2.762\times
10^{-29}J ,$ and $\omega=2000 \,{\rm rad/sec}$. This corresponds to $c=0.92$%
, $R_F=14.3 \, \mu$m, and ${\cal G}=1.23$. The difference in the densities
of the two components shows that the largest change occurs near the center
where the density is maximum.


It is to be further noted that for this case, the density distribution is
still quite cylindrical but there is a slight asymmetry, as we can see from
the graph of the difference. This asymmetry becomes more pronounced as the
interaction is increased further. In Fig. \ref{mc_chargeb} we have shown the
results of simulations with larger $a$. The density profiles were calculated
for $a=4160\,a_{{\rm Bohr}},$ $\mu_1=\mu_2=1.626\times 10^{-29} J,$ $%
\omega=6000 \,{\rm rad/sec}$. This corresponds to $c=0.98,\, R_F=3.67 \,\mu$%
m, and ${\cal G}=1.31$.

As discussed earlier, phase separation can also occur when $N_1>>N_2$. As an
illustration, we show in Fig. \ref{mc_charge2} the density profiles for
components 1 and 2 for the case $a=104 \,a_{{\rm Bohr}},$ $%
\mu_1=2.6016\times 10^{-26} J,$ $\mu_2=4.336\times 10^{-26} J$ and $%
\omega=1600000 \,{\rm rad/sec}$.

The density of component 2 is small and therefore, its noise is also
substantially higher. One can clearly see the density depletion of component
2 at the center.

\section{Dynamics}

We next turn our attention to the issue of dynamics. For the classical and
boson spinodal decompositions, the fastest unstable mode occurs at a finite
wave vector. We ask if a similar situation occurs for the fermion case. We
found that the fastest unstable mode occurs at wavevector
$q=0$ at the onset of instability. For stronger coupling, many modes with $q
\sim K_F$ decay with comparable time scales. We now describe the details of
this linear stability analysis.

The energy functional (Eq. (1)) which was approximated with a {\it local
kinetic energy} depending on the density and its derivatives is only good in
the long wavelength limit. Due to this approximation, we found that the
instability has a local character and occurs first in regions of high
density. Here we will perform a linear instability analysis in the random
phase approximation (RPA) to improve upon this local picture. The linear
susceptibility $\chi$ is defined as the response of the particle density to
an external potential $V$ which could also be $\sigma$-dependent:
\begin{equation}
\delta\rho_{\sigma}({\bf r})=\sum_{\sigma^{\prime}=1,2} \int d{\bf r}%
^{\prime}\chi_{\sigma\sigma^{\prime}}({\bf r},{\bf r}^{\prime})
V^{tot}_{\sigma^{\prime}}({\bf r}^{\prime}).  \label{susc}
\end{equation}
Here $V^{tot}$ is the total self-consistent field and is the sum of the
external field and that due to the interaction: $V_{\sigma}^{{\rm tot}}=V_{{%
\sigma}}+g\delta \rho_{{\bar{\sigma}}}.$ The bare response $%
\chi_{\sigma\sigma}$ can be obtained from the usual Lindhard expression\cite
{fetter}. Since there is no term in the Hamiltonian that interchanges the
species 1 and 2, off-diagonal terms of the susceptibility are zero ($%
\chi_{12}=\chi_{21}=0$). Taking the above into consideration, Eq. (\ref{susc}%
) can be written in the following matrix form: $\delta \rho = \chi (V + G
\delta \rho)$, leading to $\delta \rho = [1-\chi G]^{-1} \chi \, V$, where
the $2\times2$ matrix $G$ has 0 as its diagonal elements and $g$ as its
off-diagonal elements, and $\chi$ is diagonal. Consequently, an instability
will occur when the following determinant becomes zero:
\begin{equation}
{\rm Det} | {1} - {\bf {\chi}} G | = {\boldmath{1}} - g^2 \chi_{11}
\chi_{22} =0.
\end{equation}
In the case where the densities are equal, $\chi_{11} =\chi_{22}\equiv \chi$%
, the two eigenmodes are calculated as:
\begin{eqnarray}
\delta\rho_1+\delta\rho_2 &=&(1-\chi g)^{-1} \chi (V_1+V_2) \\
\delta\rho_1-\delta\rho_2 &=&(1+\chi g)^{-1} \chi (V_1-V_2).
\end{eqnarray}
The first mode corresponds to a density fluctuation, and the second mode $%
\delta\rho_1-\delta\rho_2$ represents the phase separation instability in
which we are interested. The response corresponding to this mode is given by
$\epsilon(q,w)=[1+g\chi(q,w)]$. The instability decay time $\nu^{-1}$ is
determined from the formula $\epsilon(q,i\nu)=0$, since, in this case, any
infinitesimal external potential will lead to a large change in the density.
There exists a $q=q_0$ such that $\nu(q_0)$ is largest. This determines the
spinodal wavevector of the fermionic system as it indicates the mode with
fastest growth. In what follows, we will be treating the constant external
potential problem where the Fermi momentum is ${\bar{k}}$.
For the confined case, one can consider ${\bar{k}}$ to be a local function
related to the density by ${\bar{k}}({\bf r}) = [6 \pi^2 \rho({\bf r})]^{1/3}
$. 
From the Lindhard expression\cite{fetter} for $\chi$ (real
frequencies), we obtain, after correcting for a spin degeneracy
factor of 2, the corresponding dimensionless response ${\bar
\chi}= -4\pi^2\hbar^2 \chi(q,i\nu)/m{\bar{k}}$ for imaginary
frequencies:
\begin{equation}
{\bar \chi(q,i\nu)}=1+ {\frac{1 }{2q}}(1 +({\nu/ q})^2- (q/2)^2)\, {\rm Log}[%
\frac{(1 + q/2)^2+(\nu/q)^2}{(1 -q/2)^2+(\nu/q)^2}]
\end{equation}
\[
-{\frac{\nu }{q}} \left ({\rm tan}^{-1}[{\frac{\nu/q }{(1- {q/ 2})}}] +{\rm %
tan}^{-1}[{\frac{\nu/q }{(1+ {q/ 2})}}]\right ).
\]
Here $q$ is in units of ${\bar{k}}$ and $\nu$, in units of $\hbar/2{\bar E}%
=m/ \hbar {\bar{k}}^2$. The three-dimensional plot of ${\bar \chi}$ as a
function of $q$ and $\nu$ is shown in Fig. \ref{chi}.

The equation $\epsilon(q,w)=[1\pm g\chi(q,w)]=0 $ implies that the
instability points for phase segregation with a repulsive interaction ($g>0$%
) and that of density modulation with an attractive interaction are exactly
the same within RPA. This is also in agreement with the analysis of section
II E where it was shown that "magnetic" instability occurs for repulsive
interactions, and "density wave" instability may occur for attractive
interactions. Although the susceptibility can be both negative or positive,
for a coupling of fixed sign, one should only consider the physically
correct situation. In our case, for positive $g$, only the ``magnetic"
instability, i.e. $\chi=-1/g$ should be considered.

Now since $g\chi=-{\bar \chi}\,{\bar{k}} a/\pi$, the instability
condition implies ${\bar c}{\bar \chi}=1$ where ${\bar
c}={\bar{k}} a/\pi$. The maximum of $\bar \chi$ is obtained for $q
\to 0$ and $\omega \to 0$ where it tends to 2. From this result,
we arrive at the conclusion that there is no solution to
$\epsilon(q,i\nu)=0$ for ${\bar c} < 0.5$ and no instability
develops. For larger values of ${\bar c}$, the plane $z= 1/{\bar
c}$ intersects the surface of $\bar \chi$ on a curve which is
displayed in Fig. \ref{contours}. The inverse decay time $\omega$
as a function of the wavevector in units of ${\bar{k}}$ is shown
in this figure. 
As can be seen, the fastest
unstable mode occurs at wavevector $q=0$ and $\omega=0$ at the
onset of the instability (${\bar c} = 0.5$) in agreement with Eq.
\ref {ccritical} previously derived. Indeed the instability
calculation derived in the previous section focused on the long
wavelength aspect of the problem.

For stronger couplings, many modes with $q \approx {\bar{k}}$ decay with
comparable time scales of the order of $\hbar/E_F$, but those with shortest
timescales (i.e. largest $\omega$) prevail.

In the really strong interaction limit, further phase separation can take
place either via tunnelling\cite{tunneling1,tunneling2} or via quantum
motion of the domain walls. We hope to investigate this further in the
future.

The behavior of the wavevector of instability is similar to that
of the classical spinodal decomposition, which we briefly
recapitulate here. The current $J$ can be related to the free
energy $F$ by Fick's law: $J=c\nabla F $ for some constant c.
After the onset of instability, $F=(-A+Bq^2)\delta \rho_q$. As one
goes from the onset of instability, $A$ starts to become non-zero.
In addition, there is the particle conservation equation
$-\partial_t \rho =\nabla \cdot J$. Combining the above two
equations, we obtain $i\omega \delta \rho_q=c q^2(-A+B q^2)\delta
\rho_q$. The fastest mode occurs
at a wavevector $q_c=\sqrt{ A/2B}$. Thus at the onset of instability, $q_c=0$%
. $q_c$ becomes larger as one goes away from the instability
point.

\section{Conclusion}

In conclusion we have investigated the statics and dynamics of the spatial
phase segregation process of a mixture of fermion atoms in a harmonic trap
using the density functional theory and the random phase approximation. As
the coupling starts to increase, even with the same chemical potential,
equilibrium distribution with unequal densities starts to appear, which
quite often do not exhibit axially symmetric correlations. Similar to the
classical and Bose spinodal decomposition cases, the fastest mode for the
initial phase segregation occurs at a finite wave-vector.
The condition of instability corresponds to a large interaction, which may
be achieved experimentally with the atoms close to a Feshbach resonance.

The instability calculation for the phase segregation phenomena
discussed here is related to the instability calculation for the
antiferromagnetic transition of the electron gas. In the electron
gas, this is enhanced when there is nesting of Fermi surface such
as in Cr or in one dimensional materials. The transition always
stops after the $2K_F$ instability due to the long range nature of the
Coulomb interaction, and no further ``segregation'' takes place.

An interesting situation is the one dimensional trap as it would exhibit a
much stronger instability. In mean field, the one dimensional density
difference response function $\epsilon(2K_F)=1/[1+K_F a \,{\rm Log}(T/E_F)]$
is logarithmically divergent at zero temperature. The transition temperature
occurs at $T_c=E_F e^{-1/K_Fa}.$ One dimensional trap, which can be realized
for small values of $\lambda$, has been extensively studied\cite{mit,demarko}
and we expect a higher tendency towards phase segregation in that case as
well.
\begin{acknowledgments}

\noindent S.T. Chui is partly supported by NASA under contract no.
NAG8-1427. He, KE and VK thank the Institute for Materials Research
for the kind hospitality, where the main body of this work was
completed.

\end{acknowledgments}




\begin{figure}[tbp]
\caption{Top: Dimensionless density versus dimensionless radius $\bar r/R$
for ${\cal G}=1$. One of the asymmetric solutions (A2) is depleted at the
center while the other one has a large concentration. For $\bar r/R$ larger
than 0.51 both asymmetric solutions join the symmetric density profile. The
sharp features around this point are due to the neglect of the gradient
terms. Bottom: Universal curve of rescaled density ${\cal N}=n \,{\cal G}^3$
versus rescaled distance from the border ${\cal P}=(1-{\bar r}^2/R^2)\,{\cal %
G}^2$, valid for all coupling strengths ${\cal G}$. Note that $0<{\cal P}<1$%
, and for the symmetric case ${\cal N}_{{\rm max}} = 0.43$ (${\bar r} = 0$
or ${\cal P} = 1$). }
\label{thermo}
\end{figure}


\begin{figure}[tbp]
\caption{Snap shot of the density profile at z=0 as a function of
x and y in the weak coupling limit $c=0.032$.}\label{mc_charge}
\end{figure}

\begin{figure}[tbp]
\caption{Snap shot of the density profile of components 1 (top) and 2
(bottom) at z=0 as a function of x and y in the strong coupling limit $%
c=2.39, \, \protect\omega=300$ rad/sec.}\label{mc_charge1}
\end{figure}


\begin{figure}[tbp]
\caption{Snap shot of the density profile of components 1 and 2
and their difference at z=0 as a function of x and y in the strong
coupling limit ($c= 0.92, \, \protect\omega=2000$
rad/sec).}\label{mc_chargea}
\end{figure}


\begin{figure}[tbp]
\caption{Snap shot of the density profiles of components 1 and 2 at z=0 as a
function of x and y in the strong coupling limit ($c= 0.98, \, \protect\omega%
=6000$ rad/sec).}\label{mc_chargeb}
\end{figure}

\begin{figure}[tbp]
\caption{Snap shots of the density profiles at z=0 as a function
of x and y for $c_1 = 0.98, c_2=1.27 , \, \protect\omega=1600000$
rad/sec. 
Density 2 is depleted in the central region.}
\label{mc_charge2}\end{figure}

\begin{figure}[tbp]
\caption{ Surface plot of the positive part of the reduced Lindhard
susceptibility (${\bar \protect\chi}$) as a function of $q/{\bar{k}}$ and
the imaginary frequency.}
\label{chi}
\end{figure}

\begin{figure}[tbp]
\caption{ Contour plots of the Imaginary frequency Lindhard susceptibility
indicating the inverse decay time for the phase segregation mode of wave
vector $q$ for several values of the dimensionless coupling $1/c=\protect\pi/%
{\bar{k}} a$= {0,0.3,0.65,1,1.45,1.75,1.9,1.98} starting from the
outermost line representing ${\bar \protect\chi}=0$.}
\label{contours}
\end{figure}

\epsffile{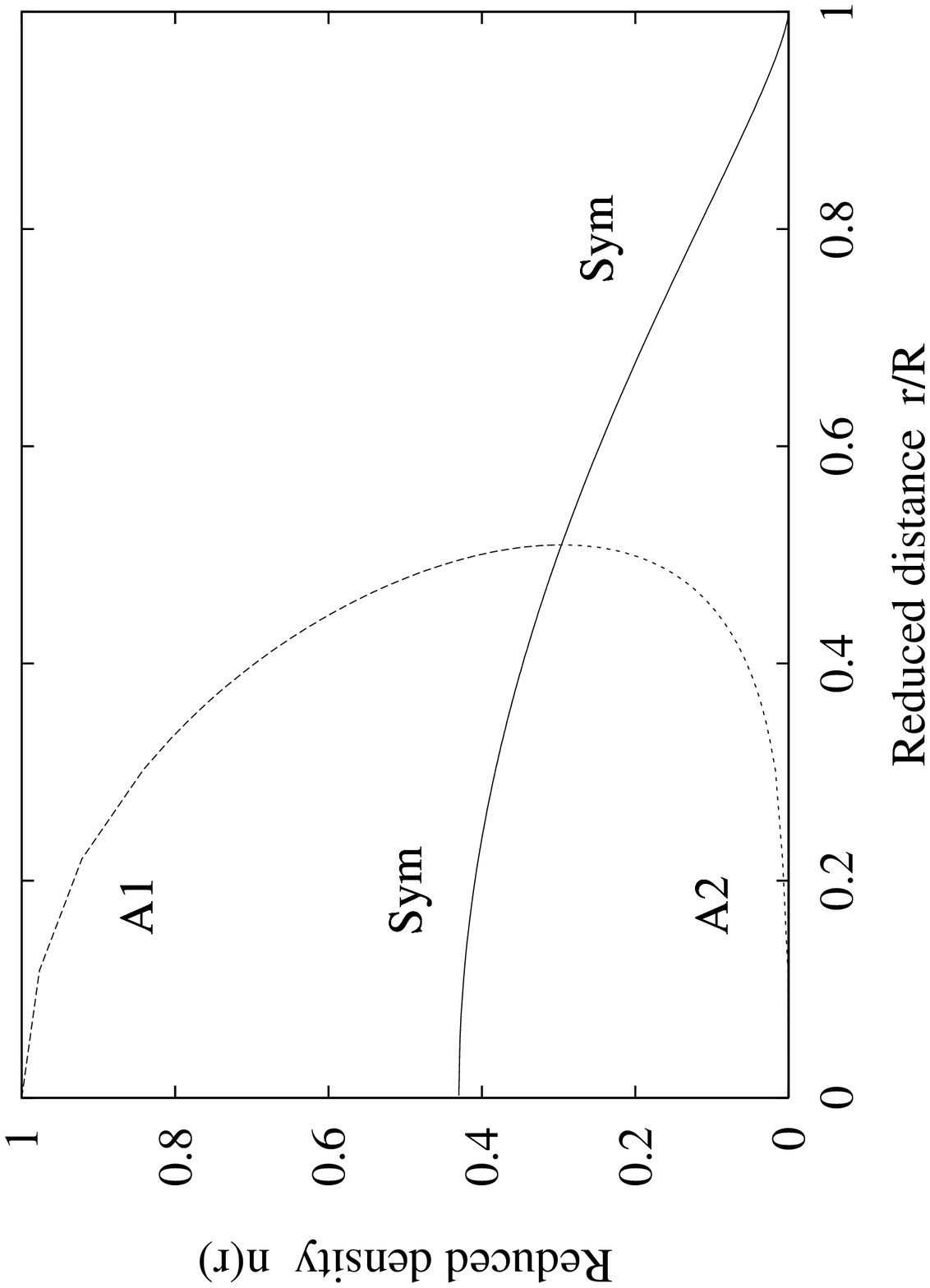}
Fig. 1 (top), K. Esfarjani et al.
\newpage

\epsffile{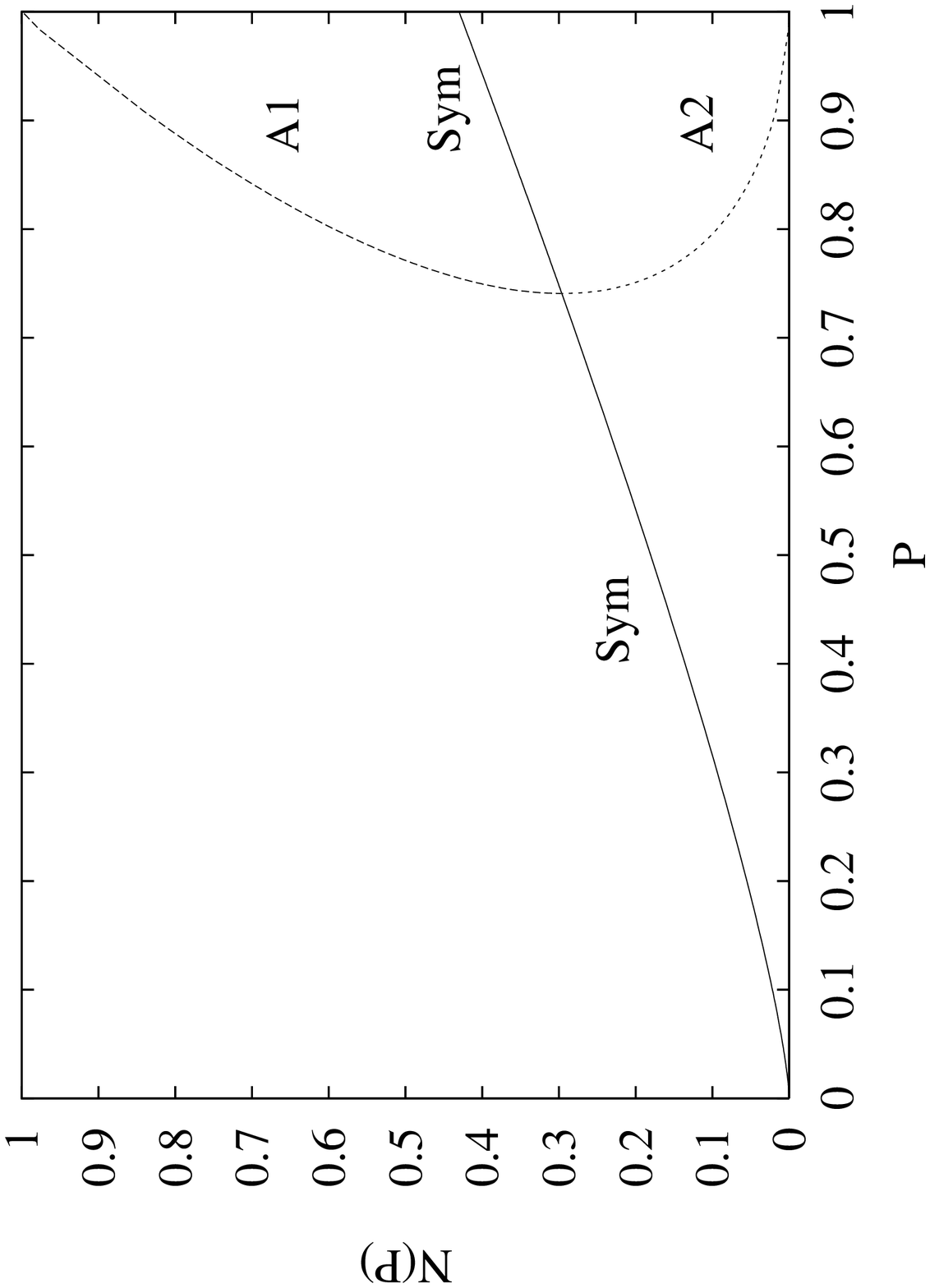}
Fig. 1 (bottom), K. Esfarjani et al.
\newpage

\epsffile{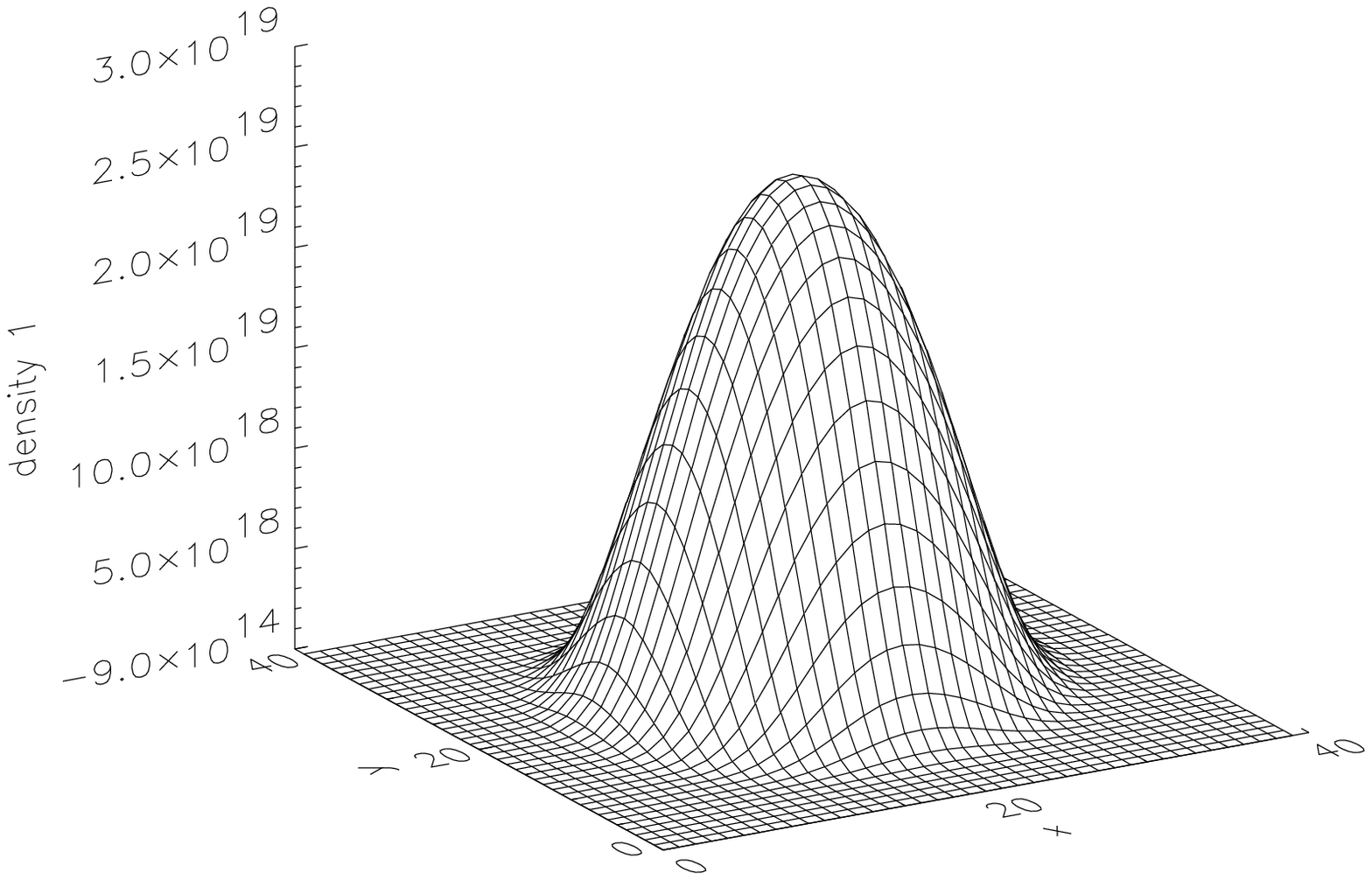}
\vspace{3 cm}
Fig. 2, K. Esfarjani et al.
\newpage

\epsffile{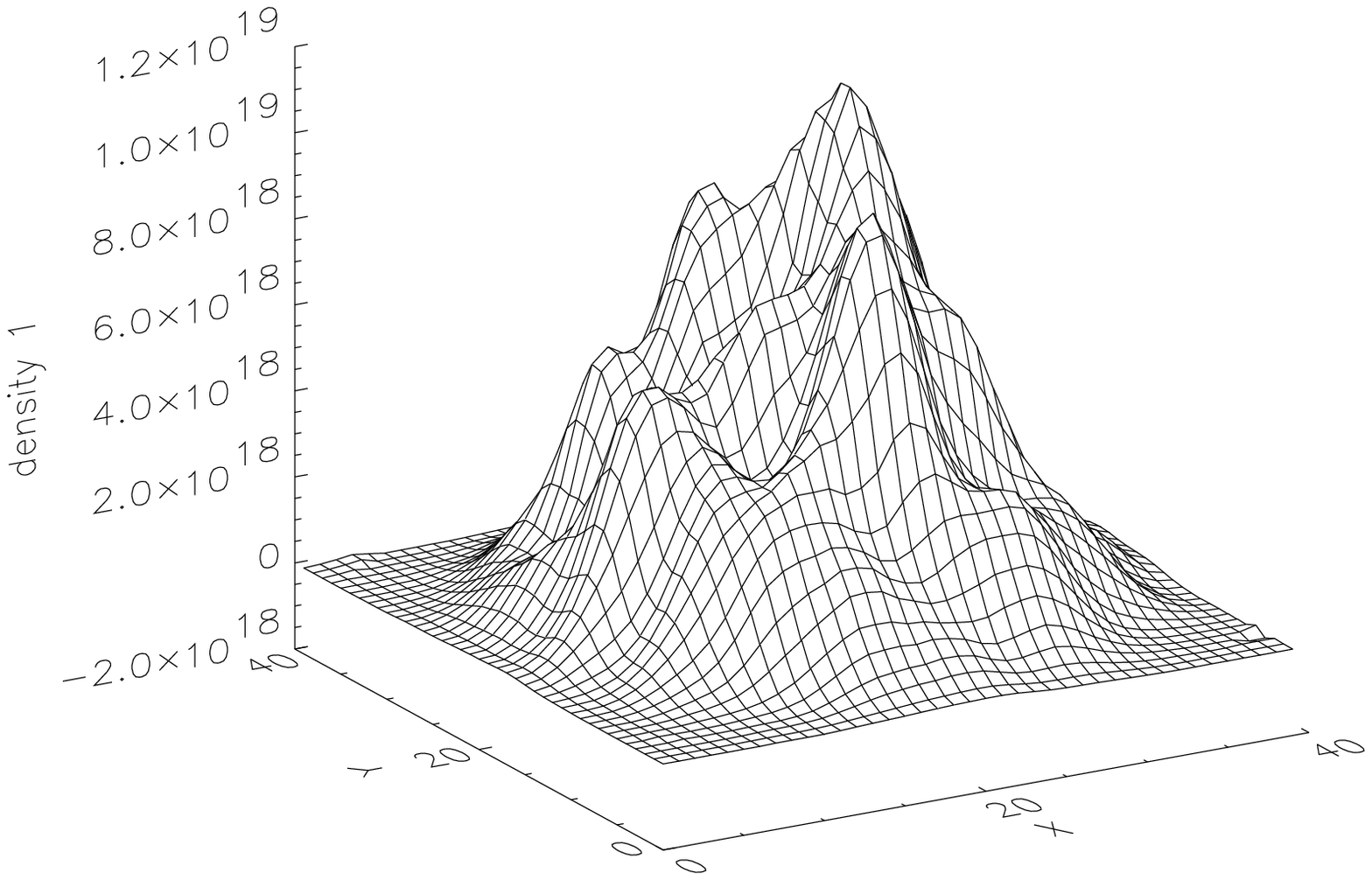}
\vspace{3 cm}
Fig. 3 (top:density 1), K. Esfarjani et al.
\newpage

\epsffile{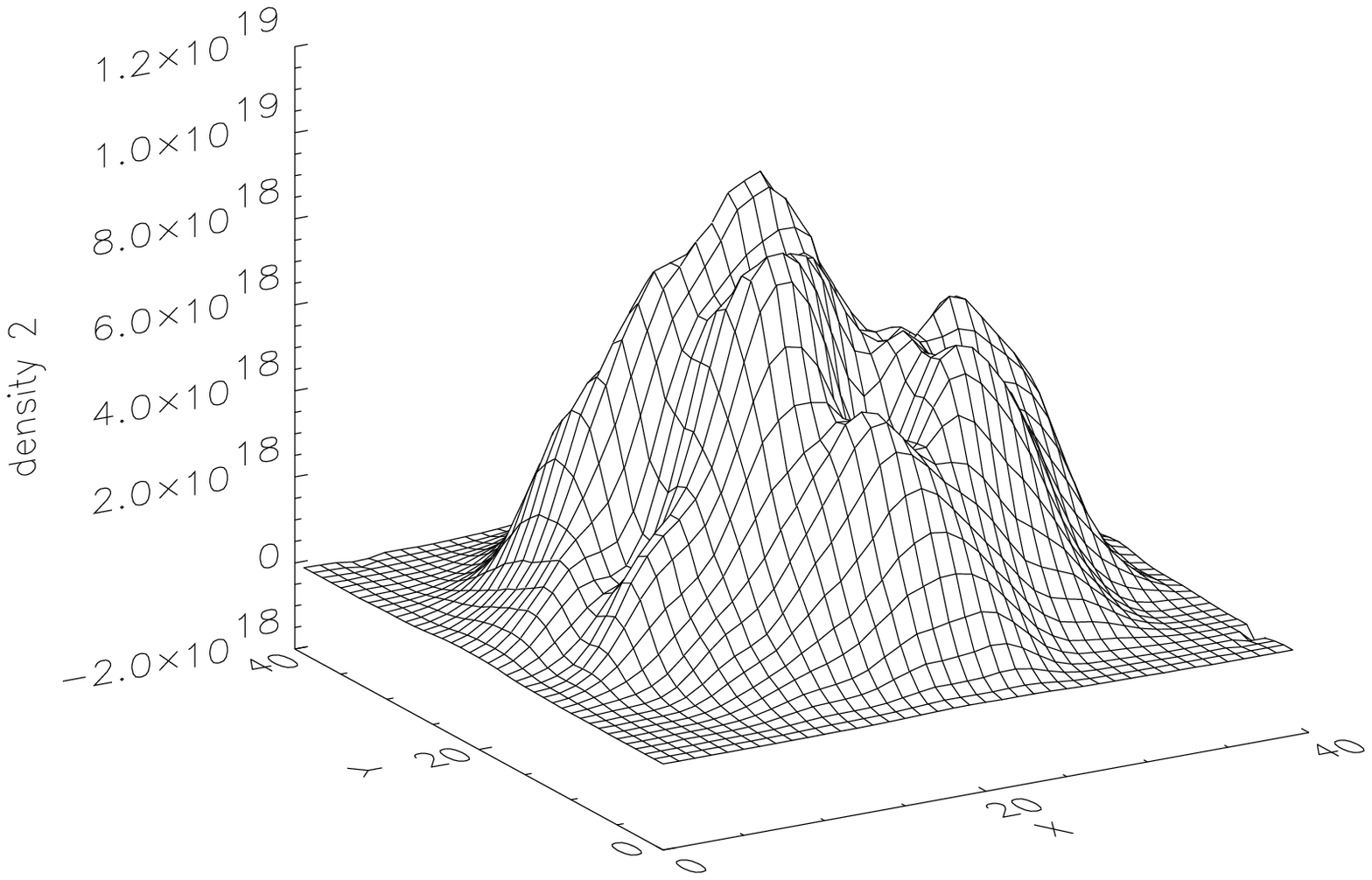}
\vspace{3 cm}
Fig. 3 (bottom:density 2), K. Esfarjani et al.
\newpage

\epsffile{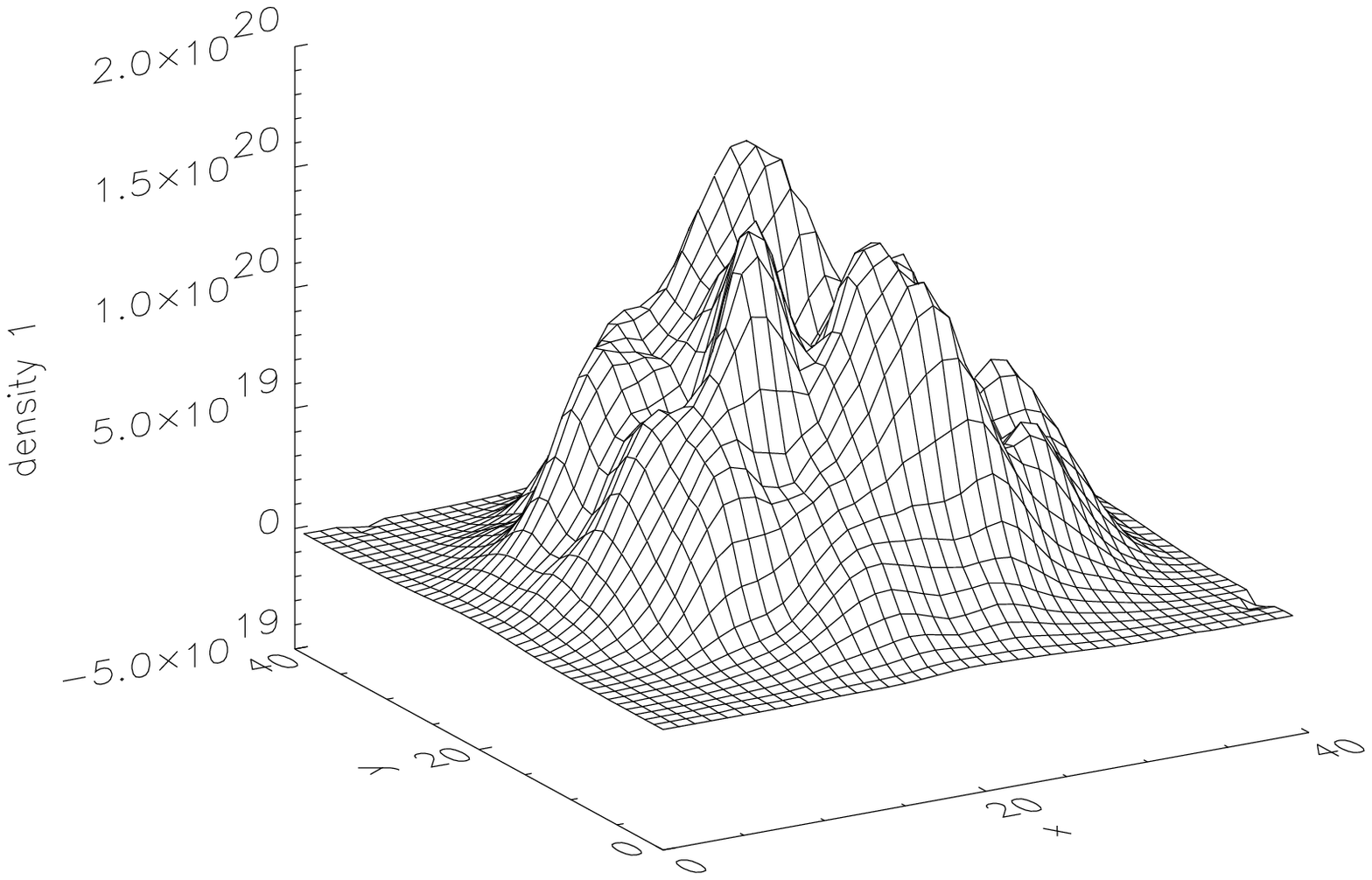}
\vspace{3 cm}
Fig. 4 (top:density 1), K. Esfarjani et al.
\newpage

\epsffile{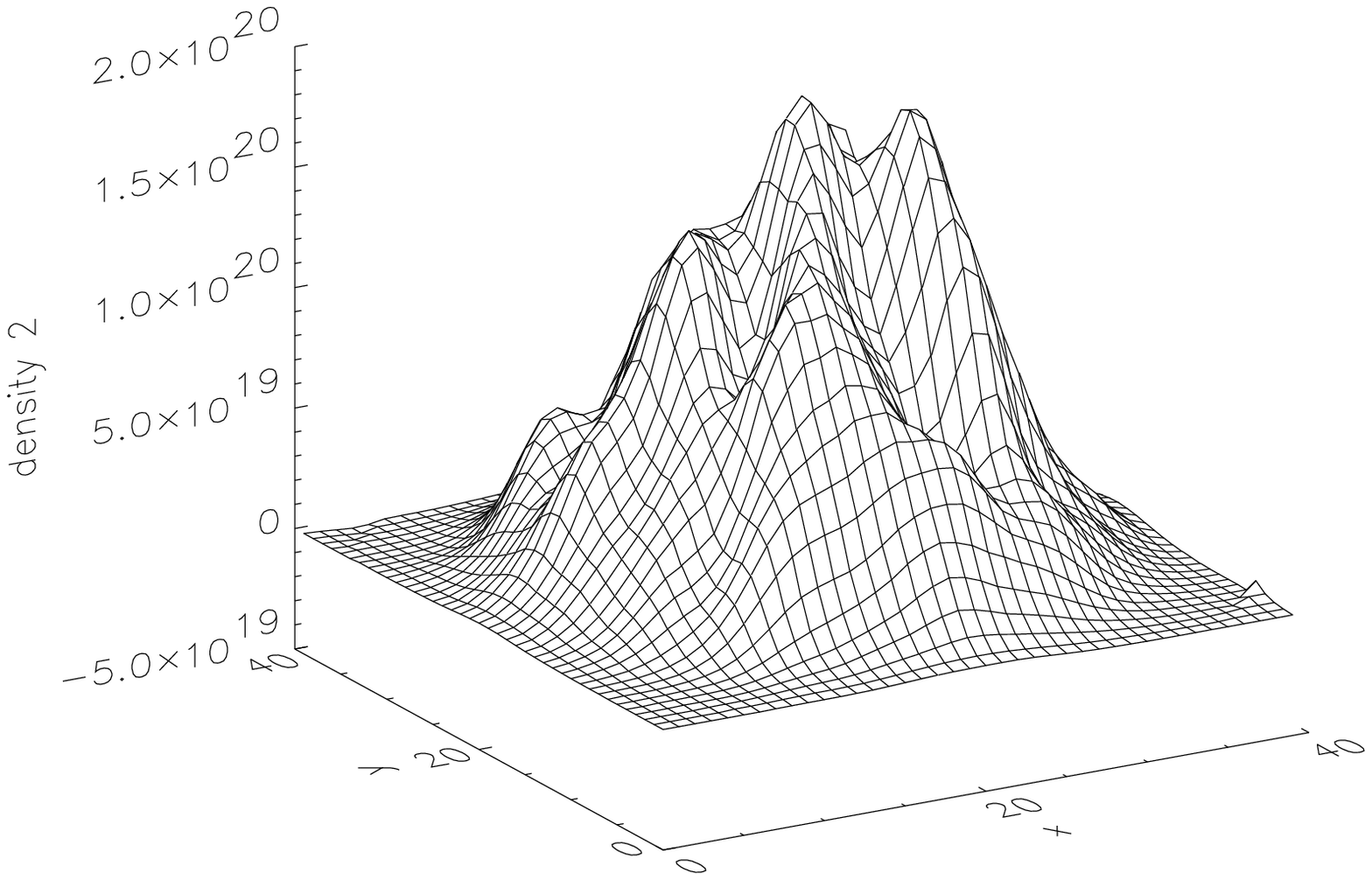}
\vspace{3 cm}
Fig. 4 (middle:density 2), K. Esfarjani et al.
\newpage

\epsffile{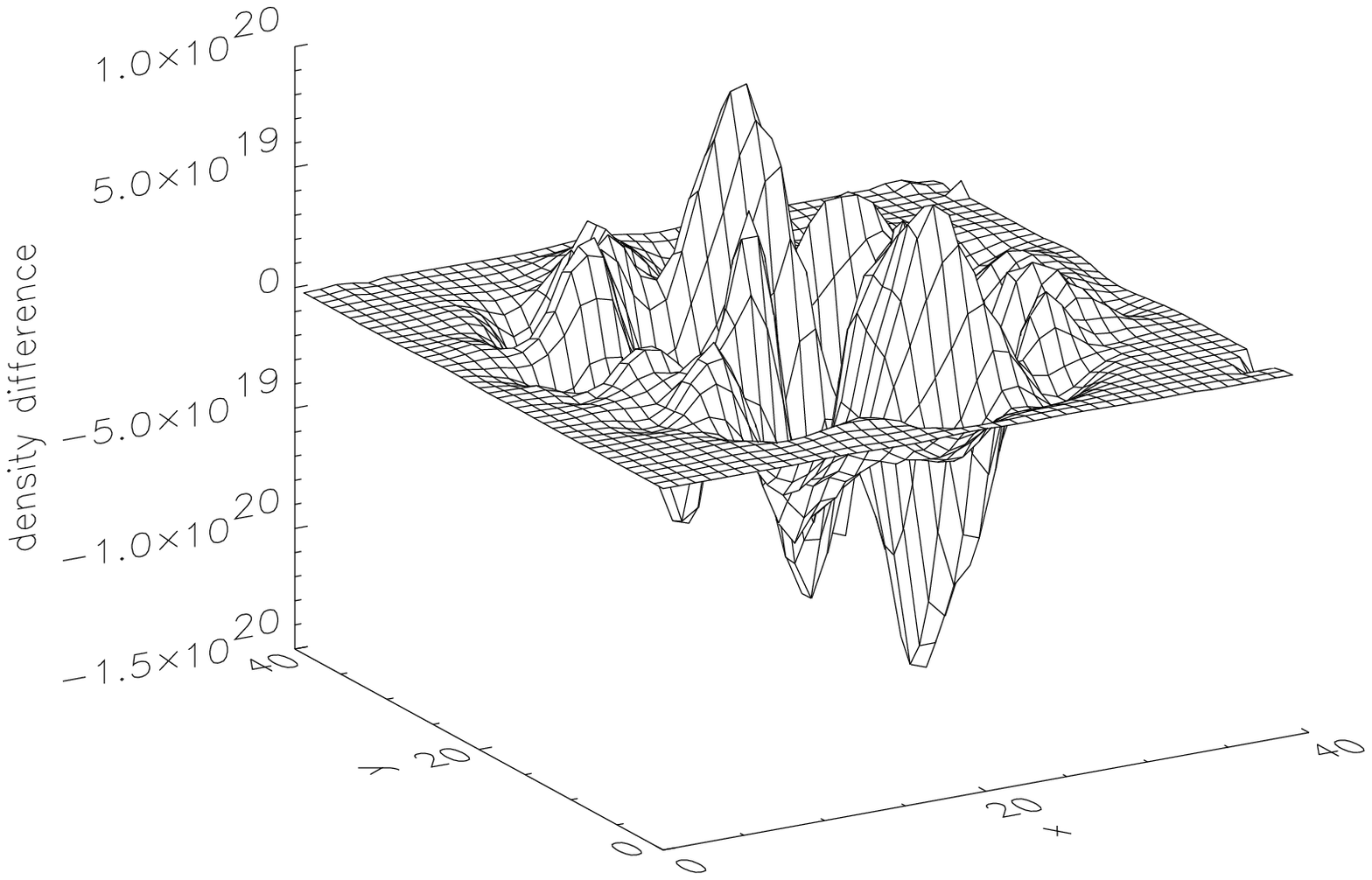}
\vspace{3 cm}
Fig. 4 (bottom:density difference), K. Esfarjani et al.
\newpage

\epsffile{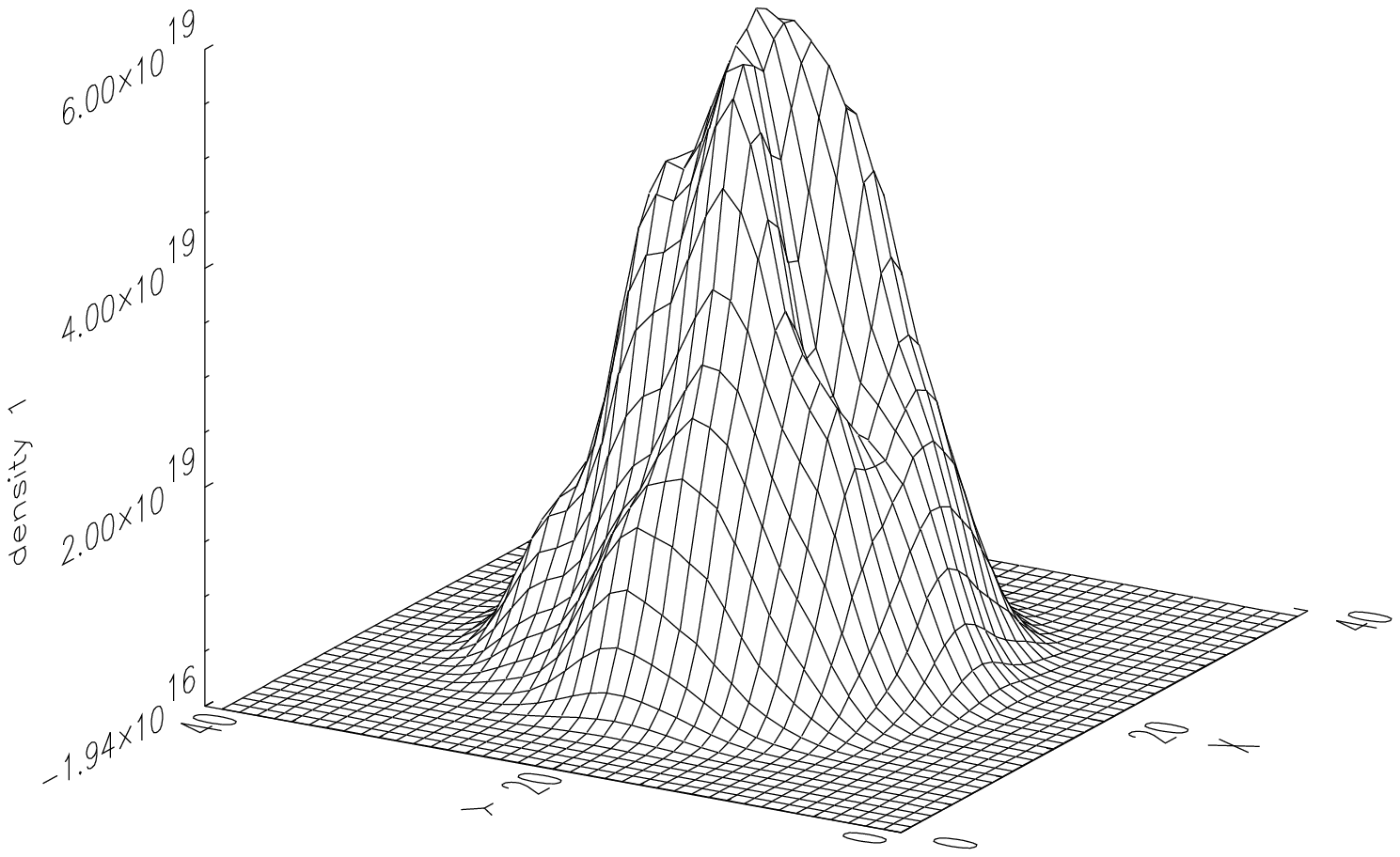}
\vspace{3 cm}
Fig. 5 (top:density 1), K. Esfarjani et al.
\newpage

\epsffile{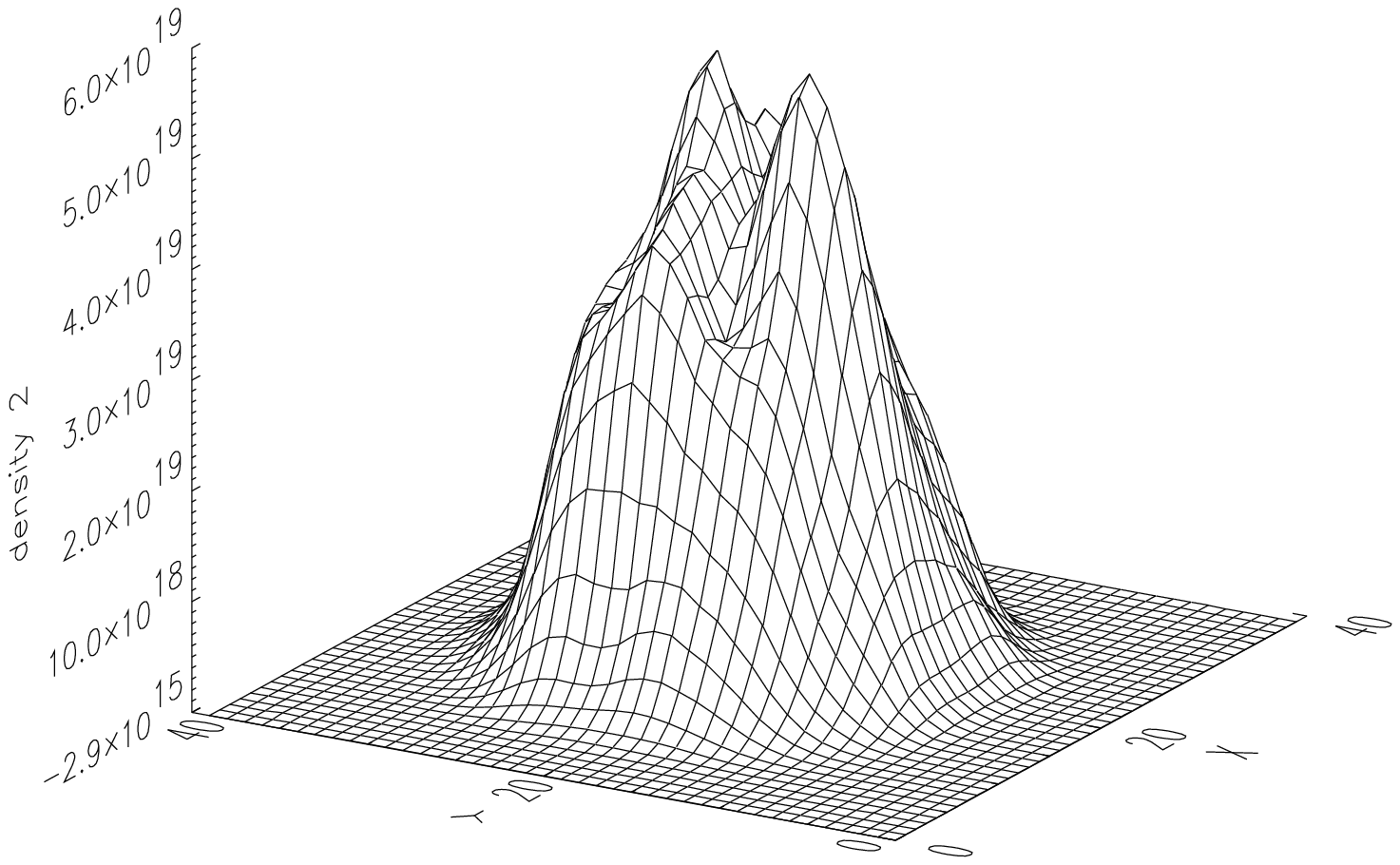}
\vspace{3 cm}
Fig. 5 (bottom:density 2), K. Esfarjani et al.
\newpage

\epsffile{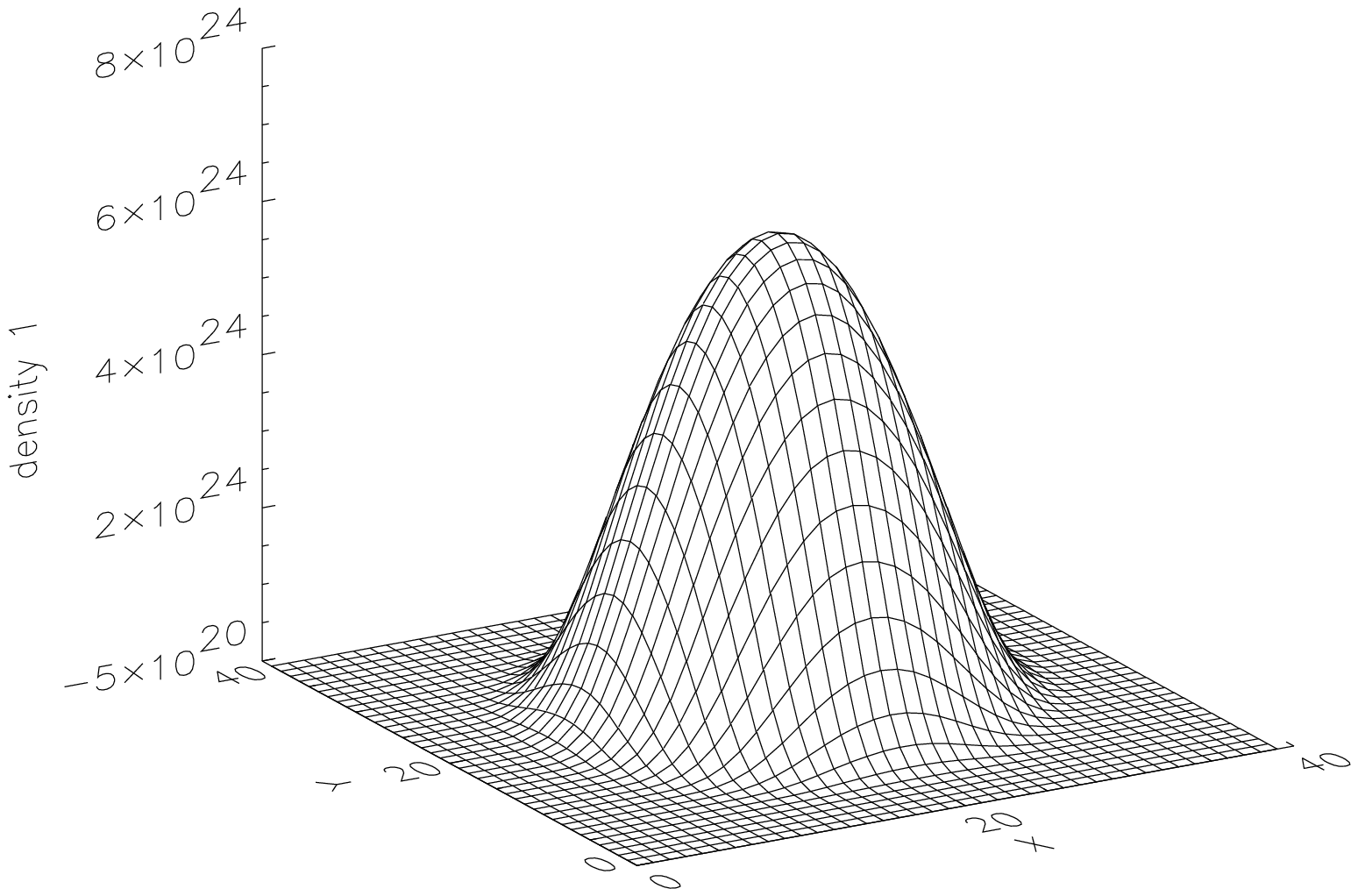}
\vspace{3 cm}
Fig. 6 (top:density 1), K. Esfarjani et al.
\newpage

\epsffile{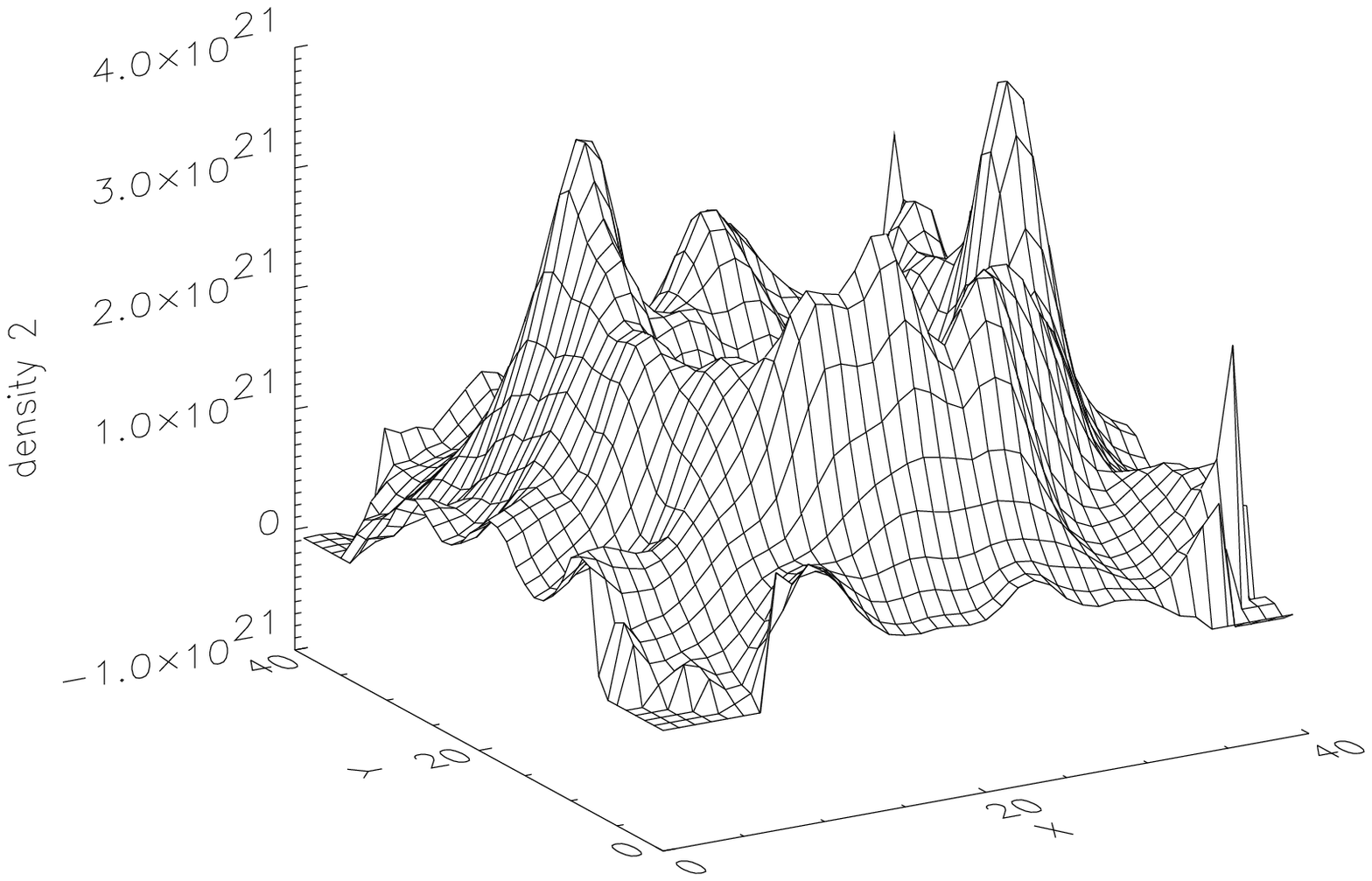}
\vspace{7 cm}
Fig. 6 (bottom:density 2), K. Esfarjani et al.
\newpage

\epsffile{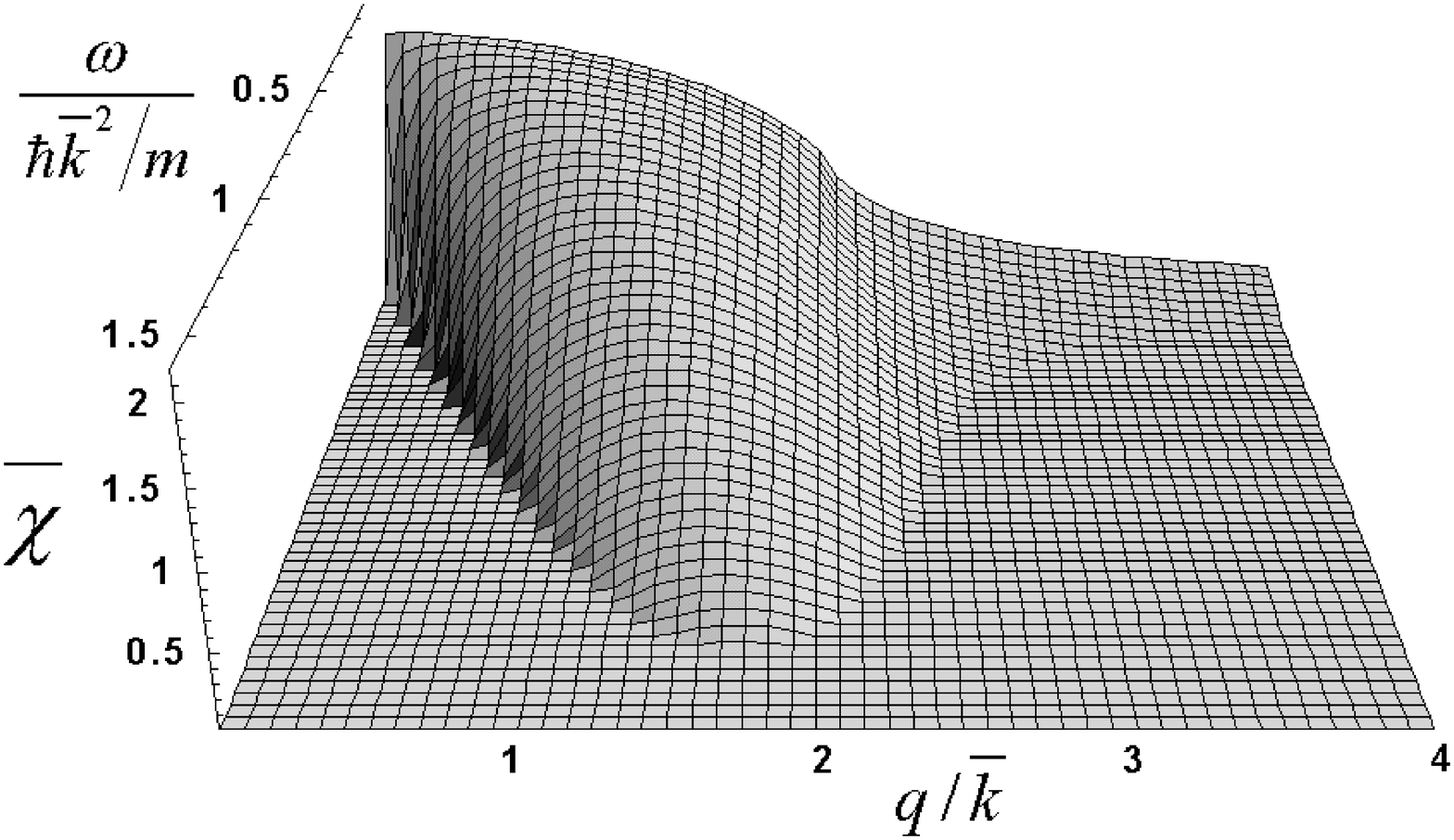}
Fig. 7 (chi bar), K. Esfarjani et al.
\newpage

\epsffile{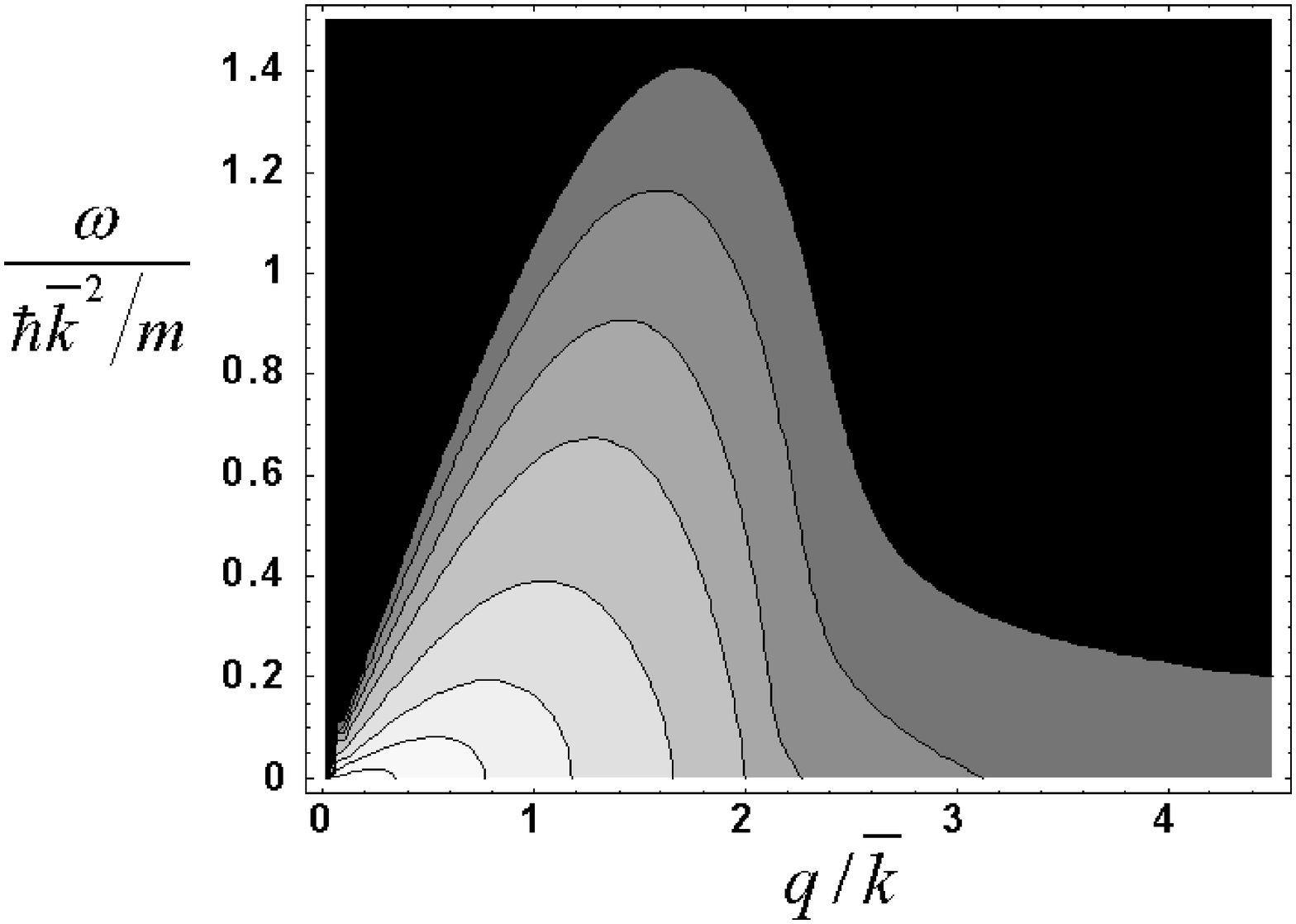}
Fig. 8 (contours of chi bar), K. Esfarjani et al.
\newpage

\end{document}